\documentclass[journal]{IEEEtran}
\ifCLASSINFOpdf
  \usepackage[pdftex]{graphicx}
  \graphicspath{{../pdf/}{../jpeg/}}
  \DeclareGraphicsExtensions{.pdf,.jpeg,.png}
\else
  \usepackage[dvips]{graphicx}
  \graphicspath{{../eps/}}
  \DeclareGraphicsExtensions{.eps}
\fi
\usepackage{epsfig}
\usepackage{booktabs}
\usepackage{epstopdf}
\usepackage{amssymb}
\usepackage{mathtools}

\usepackage{algorithmic}

\usepackage{array}

\usepackage{mdwmath}
\usepackage{mdwtab}
\usepackage{bm}
\usepackage{mathrsfs}
\usepackage{epsfig}
\usepackage{subfigure}
\usepackage{algorithmic}
\usepackage{amsmath}
\usepackage{mathrsfs}

\begin{document}

\title{Ergodic Capacity Under Power Adaption Over Fisher-Snedecor $\mathcal{F}$ Fading Channels
\thanks{Manuscript received September 20, 2018; revised November 6, 2018 and December 11, 2018; accepted January 16, 2019. The associate editor coordinating the review of this letter and approving it for publication was Dr. K. Pappi. (\emph{Corresponding author: Liang Yang.})}
\thanks{H. Zhao, A. S. Salem, and M.-S. Alouini are with the Computer, Electrical, and Mathematical Science
and Engineering (CEMSE) Division, King Abdullah University of Science and Technology (KAUST), Thuwal 23955-6900, Saudi Arabia. (email: \{hui.zhao; ahmed.salem; slim.alouini\}@kaust.edu.sa)}
\thanks{L. Yang is with the College of Computer Science and Electronic Engineering, Hunan University, Changsha 410082, China. (email: liangyang.guangzhou@gmail.com)}
\thanks{Digital Object Identifier 10.1109/LCOMM.2019.2894648}
}

\author{Hui Zhao,~\IEEEmembership{Student Member,~IEEE}, Liang Yang,~\IEEEmembership{Member,~IEEE}, \\ Ahmed S. Salem,~\IEEEmembership{Member,~IEEE}, and Mohamed-Slim Alouini,~\IEEEmembership{Fellow,~IEEE}}

\maketitle

\begin{abstract}
In this letter, we consider a communication scenario, where the transmitter adopts different power adaption methods according to the instantaneous channel state to enhance the ergodic capacity (EC) over Fisher-Snedecor $\mathcal{F}$ fading channels. We derive closed-form expressions for the EC under different power adaption methods, as well as the corresponding asymptotic EC formulas to get some insights in the high signal-to-noise ratio region. In the numerical results section, we compare the performance of different adaptive power transmission strategies,  and demonstrate the accuracy of our derived  expressions.
\end{abstract}



\section{Introduction}

Recently, the generalized-$K$ (GK) model has been proposed to approximate composite fading, where the lognormal shadowing is approximated by the K distribution \cite{Slim_TWC,Abdi}. However, the probability density function (PDF) involves the modified bessel function, which is complicated in performance analysis. To capture composite fading more accurately and develop a tractable model, \cite{Yoo} introduced the Fisher-Snedecor $\mathcal{F}$ fading, where the root-mean square value is shaped by an inverse Nakagami-$m$ distribution. Figure 2 in \cite{Yoo} shows that the Fisher-Snedecor $\mathcal{F}$ model is much better than GK model in terms of matching the tail of the empirical cumulative density function (CDF) of composite fading. Due to the fact that the tail of the empirical CDF is the main degradation (deep fading) part, the proposed Fisher-Snedecor $\mathcal{F}$ model is more practical. Another advantage of Fisher-Snedecor $\mathcal{F}$ model is that its PDF consists of only elementary functions with respect to the random variable, and is as such expected to lead to more tractable analysis than the  GK model. Based on \cite{Yoo}, \cite{Yoo2}-\cite{Zhang} investigated the Fisher-Snedecor $\mathcal{F}$ channel in depth and  derived some important metrics. \cite{Badarneh,Kong} have extended the work of \cite{Yoo} to the maximal-ratio combining technology and physical later security. However, they only considered the fixed transmit power case, i.e., optimal rate adaptation (ORA), and did not investigate the transmit power adaption according to the instantaneous channel state to enhance the EC over Fisher-Snedecor $\mathcal{F}$ channels, which is very important and practical in the real communication system design \cite{Zhao,Slim}. Although the performance of optimal power and rate algorithm (OPRA) is in general better than that of channel inversion (CI), the system complexity of CI is much lower than that of OPRA, and thus, CI is also a common power adaption method in practice.

In this letter, we consider several adaptive transmission strategies, including OPRA, ORA, CI, and truncated CI (TCI), and derive exact closed-from expressions for the EC under those power adaption methods over Fisher-Snedecor $\mathcal{F}$ fading channels, as well as corresponding asymptotic expressions for EC in the high signal-to-noise ratio (SNR) region, because the asymptotic EC (AEC) can give us some insights and design guides in communication systems.

\section{System Model}
The PDF and CDF of the SNR ($\gamma$) at the destination over Fisher-Snedecor $\mathcal{F}$ fading channels are \cite{Yoo}
\begin{align}
{f_\gamma }\left( \gamma  \right) &= \frac{{{m^m}{{\left( {{m_s}\overline \gamma  } \right)}^{{m_s}}}{\gamma ^{m - 1}}}}{{B\left( {m,{m_s}} \right){{\left( {m\gamma  + {m_s}\overline \gamma  } \right)}^{m + {m_s}}}}}, \\
{F_\gamma }\left( \gamma  \right) &= \frac{{{m^{m - 1}}{\gamma ^m}_2{F_1}\left( {m + {m_s},m;m + 1; - \frac{{m\gamma }}{{{m_s}\overline \gamma }}} \right)}}{{B\left( {m,{m_s}} \right){{\left( {{m_s}\overline \gamma } \right)}^m}}},
\end{align}
respectively, where $m$, $m_s$, and $\overline \gamma$ denote the fading severity (the number of multipath clusters), shadowing shape, and average SNR proposed in \cite{Yoo}, respectively. $B(\cdot,\cdot)$ and ${}_2 F_1 (\cdot,\cdot;\cdot;\cdot)$ represent the Beta function and hypergeometric function \cite{Gradshteyn}, respectively. By using (9) in \cite{Yoo}, i.e., the first moment of $\gamma$, when $m_s>1$, the true average SNR ($\overline \gamma_1$) can be derived in closed-form as\footnote{We can modify the model to make the true average SNR independent of $m_s$ for $m_s>1$. However, in this letter, we stick to the model in \cite{Yoo}. We provide results parameterized by $\overline \gamma$ or $\overline \gamma_1$.}
\begin{align}
{\overline \gamma  _1}=\mathbb{E}\{\gamma\} = \frac{{{m_s}\overline \gamma  }}{{{m_s} - 1}}, \quad m_s>1.
\end{align}

The EC is defined as
\begin{align}
\overline C  = \int_0^\infty  {\ln \left( {1 + \gamma } \right)} {f_\gamma }\left( \gamma  \right)d\gamma.
\end{align}
In this letter, we assume that there is a certain average transmit power ($\overline P_t$) constraint at the transmitter, and thus, the transmission must satisfy the following constraint \cite{Slim}
\begin{align}\label{constraint}
\int_0^\infty  {\frac{{{P_t}\left( \gamma  \right)}}{{{{\overline P }_t}}}} {f_\gamma }\left( \gamma  \right)d\gamma  = 1,
\end{align}
where $P_t(\gamma)$ denotes the transmit power as a function of $\gamma$.

\section{Ergodic Capacity under OPRA}
\subsection{Exact EC under OPRA}
In the OPRA case, the transmit power is adjusted according to the instantaneous channel state, given by \cite{Slim}
\begin{align}\label{transmit}
{P_t} (\gamma) = \max \left\{ {{{\overline P }_t}\left( {\frac{1}{{{\gamma _0}}} - \frac{1}{\gamma }} \right),0} \right\},
\end{align}
where ${\gamma _0} \ge 0$ denotes the cutoff point, below which no data is transmitted.
In view of (7) in \cite{Slim} and the PDF of $\gamma$, the EC under OPRA is written by
\begin{align}
&{{\overline C}_{opra}} = \frac{{{m^m}{{\left( {{m_s}\overline \gamma } \right)}^{{m_s}}}}}{{B\left( {m,{m_s}} \right)}}\int_{{\gamma _0}}^\infty  {\frac{{\ln \left( {{\gamma  \mathord{\left/
 {\vphantom {\gamma  {{\gamma _0}}}} \right.
 \kern-\nulldelimiterspace} {{\gamma _0}}}} \right){\gamma ^{m - 1}}}}{{{{\left( {m\gamma  + {m_s}\overline \gamma } \right)}^{m + {m_s}}}}}d\gamma } .
\end{align}
After some mathematical manipulations, we can derive the EC
\begin{align}\label{C_opra}
{\overline C _{opra}} = \frac{{{{\left( {\frac{{{m_s}\overline \gamma  }}{{m{\gamma _0}}}} \right)}^{{m_s}}}G_{3,3}^{1,3}\left( {\frac{{{m_s}\overline \gamma  }}{{m{\gamma _0}}}\left| {_{0, - {m_s}, - {m_s}}^{1 - {m_s},1 - {m_s},1 - m - {m_s}}} \right.} \right)}}{{\Gamma \left( m \right)\Gamma \left( {{m_s}} \right)}},
\end{align}
where $\Gamma(\cdot)$ and $G^{\cdot,\cdot}_{\cdot,\cdot}(\cdot)$ denote the Gamma function and Meijer-G function \cite{Gradshteyn}, respectively.

By substituting \eqref{transmit} into \eqref{constraint},
the corresponding power constraint condition is given by
\begin{align}
&\int_{{\gamma _0}}^\infty  {\left( {\frac{1}{{{\gamma _0}}} - \frac{1}{\gamma }} \right)} {f_\gamma }\left( \gamma  \right)d\gamma  = 1 \notag\\
&\Rightarrow \frac{1}{{{\gamma _0}}}\left[ {1 - {F_\gamma }\left( {{\gamma _0}} \right)} \right] - \int_{{\gamma _0}}^\infty  {\frac{{{f_\gamma }\left( \gamma  \right)}}{\gamma }} d\gamma  = 1,
\end{align}
where
\begin{align}
 &\int_{{\gamma _0}}^\infty  {\frac{{{f_\gamma }\left( \gamma  \right)}}{\gamma }} d\gamma= \frac{{{{\left( {{m_s}\overline \gamma  } \right)}^{{m_s}}}}}{{B\left( {m,{m_s}} \right)}}\frac{1}{{{m^{{m_s}}}}}\int_{{\gamma _0}}^\infty  {\frac{{{\gamma ^{m - 2}}}}{{{{\left( {\gamma  + \frac{{{m_s}\overline \gamma  }}{m}} \right)}^{m + {m_s}}}}}} d\gamma \notag\\
 &\mathop  = \limits^{(a)} \frac{{{{\left( {\frac{{{m_s}\overline \gamma  }}{m}} \right)}^{{m_s}}}{}_2{F_1}\left( {m + {m_s},{m_s} + 1;{m_2} + 2;\frac{{ - {m_s}\overline \gamma  }}{{m{\gamma _0}}}} \right)}}{{B\left( {m,{m_s}} \right)\left( {{m_s} + 1} \right)\gamma _0^{{m_s} + 1}}},
\end{align}
where $(a)$ follows (3.194.2) in \cite{Gradshteyn}.
Finally, the corresponding constraint condition can be written by
\begin{align}
\resizebox{.9\hsize}{!}{$
\frac{{{{\overline F }_\gamma }\left( {{\gamma _0}} \right)}}{{{\gamma _0}}} - \frac{{{{\left( {\frac{{{m_s}\overline \gamma  }}{m}} \right)}^{{m_s}}}{}_2{F_1}\left( {m + {m_s},{m_s} + 1;{m_2} + 2;\frac{{ - {m_s}\overline \gamma  }}{{m{\gamma _0}}}} \right)}}{{B\left( {m,{m_s}} \right)\left( {{m_s} + 1} \right)\gamma _0^{{m_s} + 1}}} = 1,
$}
\end{align}
where ${{{\overline F }_\gamma }\left( \cdot \right)}$ represents the complementary CDF  of $\gamma$.

Let
\begin{align}
f\left( \gamma_0  \right) = \frac{1}{{{\gamma _0}}}\left[ {1 - {F_\gamma }\left( {{\gamma _0}} \right)} \right] - \int_{{\gamma _0}}^\infty  {\frac{{{f_\gamma }\left( \gamma  \right)}}{\gamma }d\gamma }-1.
\end{align}
By using the Leibniz rule, the derivative of $f(\gamma_0)$ with respect to $\gamma_0$ can be derived by
\begin{align}
&\frac{{\partial f\left( {{\gamma _0}} \right)}}{{\partial {\gamma _0}}} =  - \frac{{1 - {F_\gamma }\left( {{\gamma _0}} \right)}}{{\gamma _0^2}} < 0.
\end{align}
Thus, $f(\gamma_0)$ is monotonically decreasing over $\gamma_0 \in \left[ {0, + \infty } \right)$. When $\gamma_0 \to 0^+$, $f(\gamma_0) \to +\infty$, and $f(\gamma_0)<0$ for $\gamma_0 \to +\infty$, so there exists unique $\gamma_0$ for $f(\gamma_0)=0$. When $\overline \gamma \to +\infty$, $f(\gamma_0)=1/\gamma_0-1$. In this case, let $f(\gamma_0)=0$, and we have $1/\gamma_0-1=0$, and thereby $\gamma_0=1$. Our numerical results shows that $\gamma_0$ increases as $\overline \gamma$ increases, so $\gamma_0$ will always lie in the interval $[0,1]$.

\subsection{Asymptotic EC under OPRA}
To derive the AEC under OPRA in high SNRs, we first express the MeijerG function in \eqref{C_opra} in the integral form, i.e.,
\begin{align}
 &G_{3,3}^{1,3}\left( {\frac{{{m_s}\overline \gamma  }}{{m{\gamma _0}}}\left| {_{0, - {m_s}, - {m_s}}^{1 - {m_s},1 - {m_s},1 - m - {m_s}}} \right.} \right) \notag\\
&=\frac{1}{2\pi i}\int_{\mathcal{L}} {\frac{{\Gamma \left( { - y} \right)\Gamma \left( {m + {m_s} + y} \right){\Gamma ^2}\left( {{m_s} + y} \right)}}{{{\Gamma ^2}\left( {1 + {m_s} + y} \right)}}{{\left( {\frac{{{m_s}\overline \gamma  }}{{m{\gamma _0}}}} \right)}^y}dy},
\end{align}
where $\mathcal{L}$ represents the path to be followed while integrating.
For ${\frac{{{m_s}\overline \gamma  }}{{m{\gamma _0}}}}>1$, the integral is $2\pi i$ times the sum of the residues at pole points. The leading term in the expansion will be determined by the residue at the double pole $y=-m_s$, i.e.,
\begin{align}\label{G_asym}
&{\rm{Re}}{{\rm{s}}_{y =  - {m_s}}} = \frac{{\Gamma \left( { - y} \right)\Gamma \left( {m + {m_s} + y} \right){\Gamma ^2}\left( {{m_s} + y} \right)}}{{{\Gamma ^2}\left( {1 + {m_s} + y} \right)}}{\left( {\frac{{{m_s}\overline \gamma  }}{{m{\gamma _0}}}} \right)^y} \notag\\
 &\approx  \Gamma \left( m \right)\Gamma \left( {{m_s}} \right){\left( {\frac{{{m_s}\overline \gamma  }}{{m{\gamma _0}}}} \right)^{ - {m_s}}}\left[ {\ln \left( {\frac{{{m_s}\overline \gamma  }}{{m{\gamma _0}}}} \right) + \psi \left( m \right) - \psi \left( {{m_s}} \right)} \right],
\end{align}
where $\psi(\cdot)$ denotes the digamma function \cite{Gradshteyn}.
By substituting \eqref{G_asym} into \eqref{C_opra}, the AEC under OPRA in high SNRs can be derived as
\begin{align}\label{triangle}
 &{\overline C _{opra}^\infty}  = \ln \overline \gamma   + \ln \left( {\frac{{{m_s}}}{{m{\gamma _0}}}} \right) + \psi \left( m \right) - \psi \left( {{m_s}} \right).
\end{align}
Besides, the AEC under OPRA in high SNRs can be also written by
\begin{align}\label{nongamma}
\overline C_{opra}^\infty= \ln \overline \gamma   + \ln \left( {\frac{{{m_s}}}{{m{}}}} \right) + \psi \left( m \right) - \psi \left( {{m_s}} \right),
\end{align}
because $\gamma_0$ goes to unity as $\overline \gamma \to \infty$, which is proved by the previous subsection. From \eqref{nongamma}, we can see that the slope with respect to $\ln \overline \gamma$ is unity, regardless of any parameter setting. By using the relationship between $\overline \gamma_1$ and $\overline \gamma$, \eqref{nongamma} can be further written as
\begin{align}\label{C_asy_g1}
&\overline C _{opra}^\infty  = \ln \left( {\frac{{\left( {{m_s} - 1} \right){{\overline \gamma  }_1}}}{{{m_s}}}} \right) + \ln \frac{{{m_s}}}{m} - \psi \left( {{m_s}} \right) + \psi \left( m \right) \notag\\
& = \ln {\overline \gamma  _1} + \ln \left( {{m_s} - 1} \right) - \psi \left( {{m_s}} \right) + \psi \left( m \right) - \ln m, \quad m_s>1.
\end{align}
It is easy to observe that $\overline C _{opra}^\infty$ is an increasing function with respect to $m$ (or $m_s$). When $m_s \to \infty$, the AEC becomes $\overline C _{opra}^\infty  = \ln \left( {{{{{\overline \gamma  }_1}} \mathord{\left/
 {\vphantom {{{{\overline \gamma  }_1}} m}} \right.
 \kern-\nulldelimiterspace} m}} \right) + \psi \left( m \right)$, which is exactly the AEC over Nakagami-$m$ channels. This shows that there is no shadowing for $m_s \to \infty$, and ${{{{{\overline \gamma  }_1}} \mathord{\left/
 {\vphantom {{{{\overline \gamma  }_1}} m}} \right.
 \kern-\nulldelimiterspace} m}}$ denotes the average SNR of each multipath cluster.

\section{Ergodic Capacity under ORA}
The transmitter cannot adjust its transmit power and just employ a constant power, i.e., $\overline P_t$, to transmit signal to the destination. The EC in this case is given by (18) in \cite{Aldalgamouni}
\begin{align}
{\overline C _{ora}} = \frac{{G_{3,3}^{2,3}\left( {\frac{{{m_s}\overline \gamma  }}{m}\left| {_{1,{m_s},0}^{1,1,1 - m}} \right.} \right)}}{{\Gamma \left( m \right)\Gamma \left( {{m_s}} \right)}}.
\end{align}

When $\overline \gamma \to \infty$, the EC can be written by
\begin{align}
 &{\overline C_{ora} ^\infty } = \frac{{{{\left( {{m_s}\overline \gamma  } \right)}^{{m_s}}}}}{{B\left( {m,{m_s}} \right)}}\frac{1}{{{m^{{m_s}}}}}\int_0^\infty   \frac{ {\ln \left( \gamma  \right)}{{\gamma ^{m - 1}}}}{{{{\left( {\gamma  + \frac{{{m_s}\overline \gamma  }}{m}} \right)}^{m + {m_s}}}}}d\gamma.
\end{align}
After some mathematical manipulations, the AEC under ORA in high SNRs is given by
\begin{align}
{\overline C_{ora} ^\infty } &= \frac{{{{\left( {{m_s}\overline \gamma  } \right)}^{{m_s}}}}}{{B\left( {m,{m_s}} \right)}}\frac{1}{{{m^{{m_s}}}}}{\left( {\frac{m}{{{m_s}\overline \gamma  }}} \right)^{{m_s}}}\frac{{\Gamma \left( m \right)\Gamma \left( {{m_s}} \right)}}{{\Gamma \left( {m + {m_s}} \right)}} \notag\\
&\hspace{2.5cm}\cdot\left[ {\ln \frac{{{m_s}\overline \gamma  }}{m} + \psi \left( m \right) - \psi \left( {{m_s}} \right)} \right] \notag\\
 &= \ln \overline \gamma   + \ln \frac{{{m_s}}}{m} + \psi \left( m \right) - \psi \left( {{m_s}} \right),
\end{align}
which is the same as \eqref{nongamma}, and this also shows that the  EC under OPRA and ORA converges in high SNRs. For $m_s>1$, the AEC under OPA can be derived as \eqref{C_asy_g1}.

\section{Ergodic Capacity under CI}
The transmitter adjusts its transmit power to maintain a fixed SNR ($\gamma_{t}$) at the destination, i.e., ${\gamma _t} = {{\gamma {P_t}(\gamma)} \mathord{\left/
 {\vphantom {{\gamma {P_t}} {{{\overline P }_t}}}} \right.
 \kern-\nulldelimiterspace} {{{\overline P }_t}}}$.
The EC is given by (46) in \cite{Slim}
\begin{align}\label{CI}
{{\overline C}_{ci}} = \ln \left( {1 + \frac{1}{{\int_0^\infty  {\left( {{1 \mathord{\left/
 {\vphantom {1 x}} \right.
 \kern-\nulldelimiterspace} x}} \right){f_\gamma }\left( x \right)dx} }}} \right).
\end{align}
When $m \le 1$, the integral in \eqref{CI} goes to infinity, and therefore, there is no closed-form for it and ${\overline C _{ci}}=0$. When $m>1$, we can drive a closed-form for the integral in \eqref{CI}
\begin{align}\label{CI_condition}
&\int_0^\infty  {\frac{{{f_\gamma }\left( x \right)}}{x}} dx = \frac{{{m^m}{{\left( {{m_s}\overline \gamma  } \right)}^{{m_s}}}}}{{B\left( {m,{m_s}} \right)}}\int_0^\infty  {\frac{{{x^{m - 2}}}}{{{{\left( {mx + {m_s}\overline \gamma  } \right)}^{m + {m_s}}}}}dx} \notag\\
 &\mathop  = \limits^{(a)} \frac{{\Gamma \left( { - 1 + m} \right)\Gamma \left( {1 + {m_s}} \right)}}{{\Gamma \left( m \right)\Gamma \left( {{m_s}} \right)}}\frac{m}{{{m_s}\overline \gamma  }}=\frac{m}{{\left( {m - 1} \right)\overline \gamma  }}, \quad m>1,
\end{align}
where $(a)$ follows (3.194.3) in \cite{Gradshteyn}.

The corresponding power constraint becomes
\begin{align}
\int_0^\infty  {\frac{{{P_t}\left( \gamma  \right)}}{{{{\overline P }_t}}}} {f_\gamma }\left( \gamma  \right)d\gamma  = 1 \Rightarrow {\gamma _t} = \frac{1}{{\int_0^\infty  {{{{f_\gamma }\left( \gamma  \right)} \mathord{\left/
 {\vphantom {{{f_\gamma }\left( \gamma  \right)} \gamma }} \right.
 \kern-\nulldelimiterspace} \gamma }d\gamma } }},
\end{align}

When $\overline \gamma \to \infty$ in the $m>1$ case, by using $\ln \left( {1 + x} \right) \approx \ln x$ for large $x$ in \eqref{CI}, the AEC under CI is
\begin{align}
&\overline C_{ci}^\infty  = \ln \overline \gamma  + \ln \left( {\frac{{m - 1}}{m}} \right), \quad m>1.
\end{align}
For $m_s>1$ and $m>1$, $\overline C _{ci}^\infty$ can be written in terms of $\overline \gamma_1$ as
\begin{align}
\overline C _{ci}^\infty  = \ln {\overline \gamma  _1} + \ln \left( {\frac{{{m_s} - 1}}{{{m_s}}}} \right) + \ln \left( {\frac{{m - 1}}{m}} \right),
\end{align}
which shows that $\overline C _{ci}^\infty$ is an increasing function with respect to $m$ (or $m_s$).

\section{Ergodic Capacity under TCI}
\subsection{Exact EC under TCI}
To avoid compensating deep fading in CI, the transmitter adopts TCI, where a cutoff point $\gamma_0$ is used to determine whether to compensate the fading, and can be also selected to achieve a specified outage probability. The corresponding EC is given by (12) in \cite{Goldsmith}
\begin{align}\label{TCI}
{{\overline C}_{tci}} = \ln \left( {1 + \frac{1}{{\int_{{\gamma _0}}^\infty  {\left( {{1 \mathord{\left/
 {\vphantom {1 x}} \right.
 \kern-\nulldelimiterspace} x}} \right){f_\gamma }\left( x \right)dx} }}} \right){{\overline F}_\gamma }\left( {{\gamma _0}} \right),
\end{align}
where
\begin{align}
&\int_{{\gamma _0}}^\infty  {\frac{{{f_\gamma }\left( x \right)}}{x}dx}  = \frac{{{m^m}{{\left( {{m_s}\overline \gamma  } \right)}^{{m_s}}}}}{{B\left( {m,{m_s}} \right)}}\int_{{\gamma _0}}^\infty  {\frac{{{x^{m - 2}}}}{{{{\left( {mx + {m_s}\overline \gamma  } \right)}^{m + {m_s}}}}}dx} \notag\\
&\mathop  = \limits^{(a)} \frac{{{{\left( {\frac{{{m_s}\overline \gamma }}{m}} \right)}^{{m_s}}}{}_2{F_1}\left( {1 + {m_s},m + {m_s};2 + {m_s}; - \frac{{{m_s}\overline \gamma }}{{m{\gamma _0}}}} \right)}}{{B\left( {m,{m_s}} \right)\left( {{m_s} + 1} \right)\gamma _0^{{m_s} + 1}}},
\end{align}
where $(a)$ follows (3.194.2) in \cite{Gradshteyn}.
\subsection{Asymptotic EC under TCI}
To derive the AEC in high SNRs, we first truncate the Taylor expansion at ${\frac{{{m_s}\overline \gamma  }}{{m{\gamma _0}}}} =\infty$ for the closed-form expression of $\int_{{\gamma _0}}^\infty  {{{{f_\gamma }\left( x \right)} \mathord{\left/
 {\vphantom {{{f_\gamma }\left( x \right)} x}} \right.
 \kern-\nulldelimiterspace} x}} dx$ up to the lowest order term, i.e.,
\begin{align}
\resizebox{.9\hsize}{!}{$
\mathop {\lim }\limits_{\overline \gamma   \to \infty } \int\limits_{{\gamma _0}}^\infty  {\frac{{{f_\gamma }\left( x \right)dx}}{x}}  \approx \frac{{\gamma _0^{ - 1}}\left( {{m_s} + 1} \right)^{-1}}{{B\left( {m,{m_s}} \right)}}
\begin{cases}
{\frac{\Gamma \left( {{m_s} + 2} \right){\Gamma \left( {m - 1} \right)}}{{\Gamma \left( {m + {m_s}} \right)}}{{\left( {\frac{{{m_s}\overline \gamma  }}{{m{\gamma _0}}}} \right)}^{ - 1}}}, & m > 1;\\
\frac{{\ln \left( {{{{m_s}\overline \gamma  } \mathord{\left/
 {\vphantom {{{m_s}\overline \gamma  } {{\gamma _0}}}} \right.
 \kern-\nulldelimiterspace} {{\gamma _0}}}} \right) + \psi \left( 1 \right) - \psi \left( {1 + {m_s}} \right)}}{{{{\left( {1 + {m_s}} \right)}^{ - 1}}\left( {{{{m_s}\overline \gamma  } \mathord{\left/
 {\vphantom {{{m_s}\overline \gamma  } {{\gamma _0}}}} \right.
 \kern-\nulldelimiterspace} {{\gamma _0}}}} \right)}}, & m=1;\\
{\frac{-\Gamma \left( {{m_s} + 2} \right)}{{\Gamma \left( {{m_s} + 1} \right)\left( {m - 1} \right)}}{{\left( {\frac{{{m_s}\overline \gamma  }}{{m{\gamma _0}}}} \right)}^{ - m}}}, & {m < 1}.
\end{cases}
$}
\end{align}
By using $\ln \left( {1 + x} \right) \approx \ln x$ for large $x$ and $\mathop {\lim }\limits_{\overline \gamma   \to \infty } {\overline F _\gamma }\left( \gamma_0  \right) \to 1$ in \eqref{TCI}, the AEC under TCI can be derived as
\begin{align}\label{TCI_a}
\resizebox{.9\hsize}{!}{$
\overline C _{tci}^\infty  =
\begin{cases}
{\ln \overline \gamma   + \ln \left( {\frac{m-1}{{m }}} \right)}, & {m > 1};\\
\ln \overline \gamma   + \ln \left( {\frac{{B\left( {1,{m_s}} \right){m_s}}}{{\ln \overline \gamma   + \ln \left( {{{{m_s}} \mathord{\left/
 {\vphantom {{{m_s}} {{\gamma _0}}}} \right.
 \kern-\nulldelimiterspace} {{\gamma _0}}}} \right) + \psi \left( 1 \right) - \psi \left( {1 + {m_s}} \right)}}} \right), & m=1;\\
{m\ln \overline \gamma  + \ln \left( {\frac{{B\left( {m,{m_s}} \right)\left( {1 - m} \right)}}{{\gamma _0^{ - 1 + m}{{\left( {{m \mathord{\left/
 {\vphantom {m {{m_s}}}} \right.
 \kern-\nulldelimiterspace} {{m_s}}}} \right)}^m}}}} \right)}, & {m < 1},
\end{cases}
$}
\end{align}
which shows that the slope with respect to $\ln \overline \gamma$ is unity for $m > 1$, and $m$ for $m<1$, while the AEC is not a line function with respect to $\ln \overline \gamma$ for $m=1$. Moreover, the AEC under TCI is the same as that under CI for $m>1$,
regardless of the cutoff value of $\gamma_0$.

For $m_s>1$, the AEC under TCI can be further written as
\begin{align}
\resizebox{.9\hsize}{!}
{$\overline C _{ci}^\infty  = \begin{cases}
{\ln {{\overline \gamma  }_1} + \ln \left( {\frac{{{m_s} - 1}}{{{m_s}}}} \right) + \ln \left( {\frac{{m - 1}}{m}} \right),}&{m > 1;}\\
{\ln {{\overline \gamma  }_1} + \ln \left( {\frac{{\left( {{m_s} - 1} \right)B\left( {1,{m_s}} \right){m_s}}}{{{m_s}\left( {\ln {{\overline \gamma  }_1} + \ln \left( {{{\left( {{m_s} - 1} \right)} \mathord{\left/
 {\vphantom {{\left( {{m_s} - 1} \right)} {{\gamma _0}}}} \right.
 \kern-\nulldelimiterspace} {{\gamma _0}}}} \right) + \psi \left( 1 \right) - \psi \left( {1 + {m_s}} \right)} \right)}}} \right),}&{m = 1;}\\
{m\ln {{\overline \gamma  }_1} + m\ln \left( {\frac{{{m_s} - 1}}{{{m_s}}}} \right) + \ln \left( {\frac{{B\left( {m,{m_s}} \right)\left( {1 - m} \right)}}{{\gamma _0^{ - 1 + m}{{\left( {{m \mathord{\left/
 {\vphantom {m {{m_s}}}} \right.
 \kern-\nulldelimiterspace} {{m_s}}}} \right)}^m}}}} \right),}&{m < 1}.
\end{cases} $}
\end{align}
\section{Numerical Results}
Fig. 1 plots the EC under TCI versus $\overline \gamma$ with different $m$,  where it is easy to see that the slope in the $m>1$ case is unity, while the figure for $m<1$ is $m$. Although the AEC under TCI is not a line function with respect to $\ln \overline \gamma$ for $m=1$, the slope  changes very slowly in high SNRs.   There is an increasing trend when $m$ grows, due to the increase in the number of multipath clusters.
\begin{figure}
\vspace{-0.3cm}  
\setlength{\abovecaptionskip}{-0.2cm}   
\setlength{\belowcaptionskip}{-3cm}   
  \centering
  \includegraphics[width=3 in]{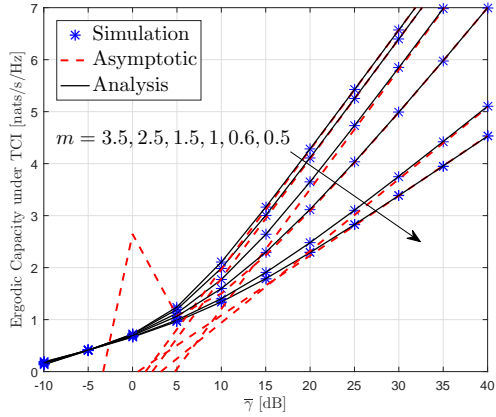}
  \caption{EC under TCI versus $\overline \gamma$ for $\overline P_t=0$ dB, $\gamma_0=0.5$ and $m_s=2.5$}
\end{figure}

In Fig. 2, we provide the EC versus the true average SNR ($\overline \gamma_1$). The EC is reduced when $m_s$ decreases, because the shadowing becomes more severe (The shadowing vanishes as $m_s \to \infty$). The slope of the EC in Fig. 2 is fixed for different $m_s$, because the slope is always unity for $m>1$.
\begin{figure}
\vspace{-0.3cm}  
\setlength{\abovecaptionskip}{-0.2cm}   
\setlength{\belowcaptionskip}{-3cm}   
  \centering
  \includegraphics[width=3.5 in]{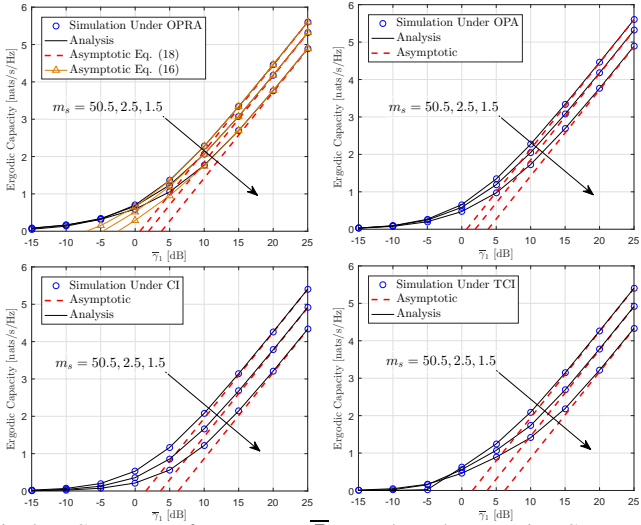}
  \caption{EC versus $\overline \gamma_1$ for $m=3.5$, $\overline P_t=0$ dB and $\gamma_0=1$ in TCI.}
\end{figure}
\begin{figure}
\vspace{-0.3cm}  
\setlength{\abovecaptionskip}{-0.2cm}   
\setlength{\belowcaptionskip}{-3cm}   
  \centering
  \includegraphics[width=3 in]{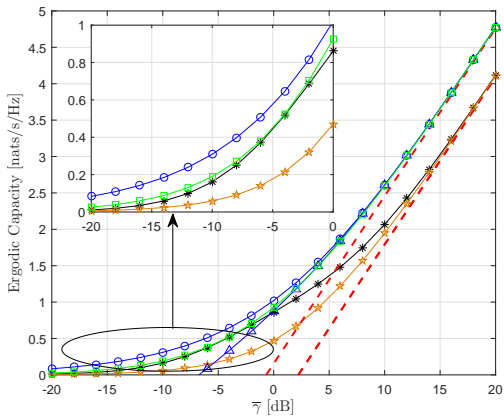}
  \caption{EC versus $\overline \gamma$ for $m_s=1.5$, $m=2.5$, $\overline P_t=0$ dB and $\gamma_0=1$ in TCI, where the circle (square, star, pentagram) symbol represents the numerical result of EC under OPRA (ORA, TCI, CI),  solid and dash lines represent corresponding analytical and asymptotic results, respectively, and the triangle line represents the asymptotic result for OPRA derived by  \eqref{triangle}.}
\end{figure}

As shown in Fig. 3,  the EC under OPRA is largest among those adaptive transmission strategies, followed by the figures for ORA and TCI, while CI is the worst case in terms of EC. The EC under TCI and CI converges in the high SNR region, because the asymptotic expressions for those two cases are the same for $m>1$, which is also the reason of the convergence of  the EC under OPRA and ORA in high SNRs. Besides, in the OPRA case,  asymptotic results from \eqref{triangle} is closer to exact results than results derived from \eqref{nongamma} in the low and medium SNR region, due to the fact that $\gamma_0$ is not close to unity until $\overline \gamma$ is large sufficiently.

\section{Conclusion}
The exact and corresponding asymptotic expressions for the EC under different power adaption schemes were derived. From asymptotic expressions,  the slope of the EC under OPRA, ORA, CI ($m>1$) with respect to $\ln \overline \gamma$ is always unity in high SNRs, regardless of any parameter setting. In contrast, the slope under TCI depends on $m$, i.e., unity for $m>1$ and $m$ for $m<1$, and the EC under TCI is not a line function with respect to $\ln \overline \gamma$ in the $m=1$ case, although the slope changes very slowly in high SNRs. Further, the EC is an increasing function with respect to $m$ (or $m_s$) in high SNRs.
From  numerical results, the performance of OPRA was best among those power adaption schemes.


\begin{thebibliography}{1}


\bibitem{Slim_TWC}
A. Laourine,  M.-S. Alouini, S. Affes, and A. Stephenne, ``On the capacity of generalized-$K$ fading channels," \emph{IEEE Trans. Wireless Commun.}, vol. 7, no. 7, pp. 2441-2445, Jul. 2008.

\bibitem{Abdi}
A. Abdi and M. Kaveh, ``K distribution: An appropriate substitute for
Rayleigh-lognormal distribution in fading-shadowing wireless channels,"
\emph{Electron. Lett.}, vol. 34, no. 9, pp. 851-852, Apr. 1998.


\bibitem{Yoo}
S. Ki Yoo, S. L. Cotton, P. C. Sofotasios, M. Matthaiou, M. Valkama, and G. K. Karagiannidis, ``The Fisher-Snedecor $\mathcal{F}$ distribution: A simple and accurate composite fading model," \emph{IEEE Commun. Lett.}, vol. 21, no. 7, pp. 1661-1664, Jul. 2017.

\bibitem{Yoo2}
S. Ki Yoo, S. L. Cotton, P. C. Sofotasios, S. Muhaidat, O. S. Badarneh, and G. K. Karagiannidis, ``Entropy and energy detection-based spectrum sensing over $\mathcal{F}$ composite fading channels," \emph{arXiv:1807.06112v1}, Jul. 2018.

\bibitem{Almehmadi}
F. S. Almehmadi, and O. S. Badarneh, ``On the effective capacity of Fisher-Snedecor $\mathcal{F}$ fading channels," \emph{Electron. Lett.}, vol. 54, no. 18, pp. 1068-1070, Sep. 2018.

\bibitem{Aldalgamouni}
T. Aldalgamouni, M. C. Ilter, O. S. Badarneh, and H. Yanikomeroglu, ``Performance analysis of Fisher-Snedecor $\mathcal{F}$ composite fading channels," in \emph{Proc. IEEE MENACOMM}, Apr. 2018, pp. 1-5.

\bibitem{Zhang}
S. Chen, J. Zhang, G. K. Karagiannidis, and B. Ai, ``Effective rate of MISO systems over Fisher-Snedecor $\mathcal{F}$ fading channels," \emph{IEEE Commun. Lett.}, vol. 22, no. 12, pp. 2619-2622, Dec. 2018.

\bibitem{Badarneh}
O. S. Badarneh, D. B. da Costa, P. C. Sofotasios, S. Muhaidat, and S. L. Cotton, ``On the sum of Fisher-Snedecor $\mathcal{F}$ variates and its application to maximal-ratio combining,"  \emph{IEEE Wireless Commun. Lett.}, vol. 7, no. 6, pp. 966-969, Dec. 2018.

\bibitem{Kong}
L. Kong, and G. Kaddoum, ``Physical layer security over the Fisher-Snedecor $\mathcal{F}$ wiretap fading channels," \emph{IEEE Access}, vol. 6, pp. 39466-39472, Jul. 2018.

\bibitem{Zhao}
H. Zhao, Z. Liu, and M.-S. Alouini, ``Different power adaption methods on fluctuating two-ray fading channels," \emph{IEEE Wireless Commun. Lett.}, to be published, DOI: 10.1109/LWC.2018.2881158.

\bibitem{Slim}
M.-S. Alouini, and A. J. Goldsmith, ``Capacity of Rayleigh fading channels
under different adaptive transmission and diversity-combining techniques," \emph{IEEE Trans. Veh. Technol.}, vol. 48, no. 4, pp. 1165-1181, Jul. 1999.


\bibitem{Gradshteyn}
I. S. Gradshteyn, I. M. Ryzhik, \emph{Table of Integrals, Series, and Products}, 7th edition. Academic Press, 2007.

\bibitem{Goldsmith}
A. J. Goldsmith, and P. P. Varaiya, ``Capacity of fading channel with
channel side information," \emph{IEEE Trans. Inf. Theory}, vol. 43, no. 6,
pp. 1986-1992, Nov. 1997.



\end{thebibliography}
\end{document}